\documentclass[epj]{svjour}
\usepackage{graphics}
\begin{document}
\title{Probing the nucleon structure with CLAS}
\subtitle{\ Highlights of recent results.}
\author{Volker D. Burkert, for the CLAS collaboration.}
%
%
\institute{Jefferson Lab, Newport News, Virginia, USA}
\date\today
%
\abstract{An overview of recent results with CLAS is presented with emphasis on the nucleon resonance program and related topics.
} 
\maketitle
\section{Introduction}
\label{intro}
The beauty of the electromagnetic probe is that it allows us to 
efficiently address the central question of hadron physics: What are the 
relevant degrees of freedom at varying distance scales?  Using electron 
beams we can vary the space-time resolution and momentum transfer to the 
nucleon independently. In doing so we probe the effective degrees of 
freedom in the nucleon from hadrons, constituent quarks, to 
elementary quarks and gluons. The study of nucleon resonance transitions, which 
is the focus of this workshop, provides a testing ground for our 
understanding of these effective degrees of freedom. Using the $SU(6) 
\otimes O(3)$ classification scheme of the symmetric constituent quark 
model (CQM), the known states can be sorted into supermultiplets of energy 
and orbital angular momentum of the 3-quark system. In this talk I will 
highlight some of the new CLAS results for the $N\Delta(1232)$ 
transition, and for some of the higher excited states of the nucleon. 
These data allow us to address questions about 
the underlying degrees of freedom of some of the well known states such 
as the Roper $P_{11}(1440)$, and $S_{11}(1535)$, both of which have also 
been presented using non-quark degrees of freedom. Studying the 
resonance transitions will allow us to make more definite 
statements about the nature of these states. Then I will discuss the 
well known problem of the "missing states", i.e. resonances predicted 
within the $SU(6) \otimes O(3)$ symmetry. Specific mass ranges are
predicted in explicit models that break degeneracies through spin-spin interactions. However, many of these states 
have not been identified in experimental analysis. Going somewhat above 
the nucleon resonance region, there is also new information on the spin 
structure of the nucleon and the resulting effects on the parton 
distribution function from recent very precise CLAS data. Finally, I 
will briefly discuss the first DVCS results that cover a broad 
kinematics regime, and what we can learn from them about the generalized 
parton distributions (GPDs).

\begin{figure}
\resizebox{0.45\textwidth}{!}{%
  \includegraphics{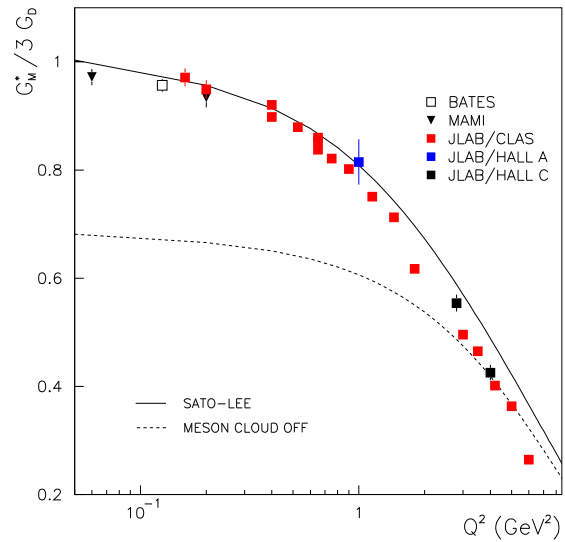}}
\caption{Magnetic form factor for the $N\Delta$ transition.}
\label{fig:G*m}       
\end{figure}
\section{The $N\Delta(1232)$ transition}
\label{sect:1}
The $N\Delta(1232)$ transition has been studied for more than 50 years 
with various probes. But only in the past decade have the experimental 
tools in electron scattering become available that 
enabled precise determinations of the magnetic transition 
form factor in $\pi^0$ production from protons with photon virtualities 
up to $Q^2=6$~GeV$^2$. Benchmark results from JLab~\cite{joo2002,ungaro2006,kelly2005,frolov1999}, 
MIT-Bates~\cite{sparveris2003}, and MAMI~\cite{sparveris2007} are shown in Fig.~\ref{fig:G*m} 
relative to the dipole form which approximately describes the elastic magnetic 
form factor of the proton. The theoretical description in the Sato-Lee 
dynamical model includes dynamical pion contributions that are needed to 
explain the magnitude of $G^*_M$ especially at lower $Q^2$. It is found that the 
pion contributions make up more than 30\% of the total amplitude at the 
photon point, and remain sizeable even at the highest $Q^2$.   

\begin{figure}
\resizebox{0.52\textwidth}{!}{%
  \includegraphics{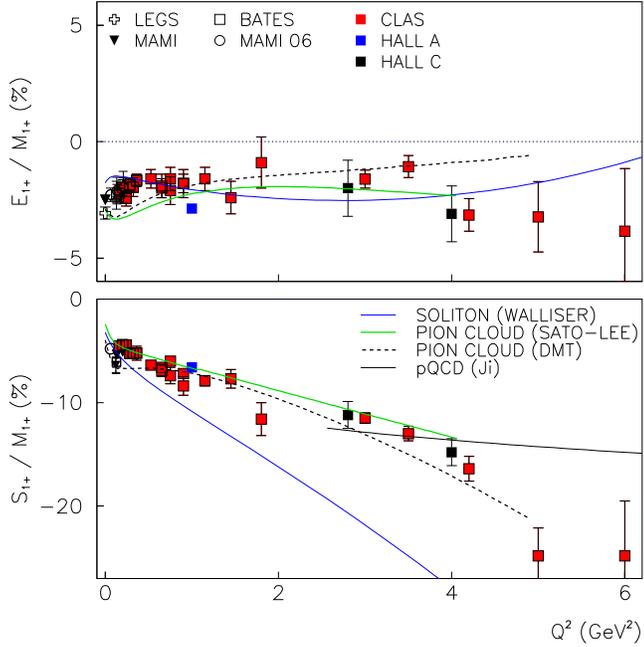}}
\caption{The electric and scalar quadrupole ratios $R_{EM}$ and $R_{SM}$ for the $N\Delta$ transition.}
\label{fig:remrsm}       
\end{figure}

The electric and scalar quadrupole contributions, expressed as fractions of the magnetic 
dipole transition and given by the ratio $R_{EM}=Im(E_{1+})/Im(M_{1+})$
and the ratio $R_{SM}=Im(S_{1+})/Im(M_{1+})$, which 
are both shown in Fig.~\ref{fig:remrsm}. $R_{EM}$ remains small and negative 
even at the highest $Q^2$, in the range from -2\%  to -4\%, and shows no  
indication of a trend towards the predicted asymptotic behavior of $R_{EM} \rightarrow 
+100\%$ at $Q^2 \rightarrow \infty$. Although $R_{SM}$ shows a different behavior, and rises in 
magnitude with $Q^2$, it also shows no indication of approaching the predicted asymptotic 
behavior, $R_{SM}\rightarrow~constant$ for 
$Q^2\rightarrow \infty$. Both of these results present serious 
challenges to theory. One may expect that Lattice QCD (LQCD) will soon be 
able to calculate these ratios accurately up to high $Q^2$. First 
computations in quenched LQCD~\cite{alexandrou} have produced results at 
lower $Q^2$ that compare favorably with the measured $R_{EM}$ values. 
However, they also reveal shortcomings for $R_{SM}$ at the lowest $Q^2$ 
values where pion contributions are expected to be important and may be underestimated
in quenched QCD.

\section{The second resonance region}
\label{sect:2}
In the mass region above the $\Delta(1232)$ there are 3 excited nucleon 
states, the Roper $P_{11}(1440)$, the $S_{11}(1535)$ and the 
$D_{13}(1520)$. Each of these resonances has features that makes 
their investigation particularly interesting.

\begin{figure}
\resizebox{0.4\textwidth}{!}{%
  \includegraphics{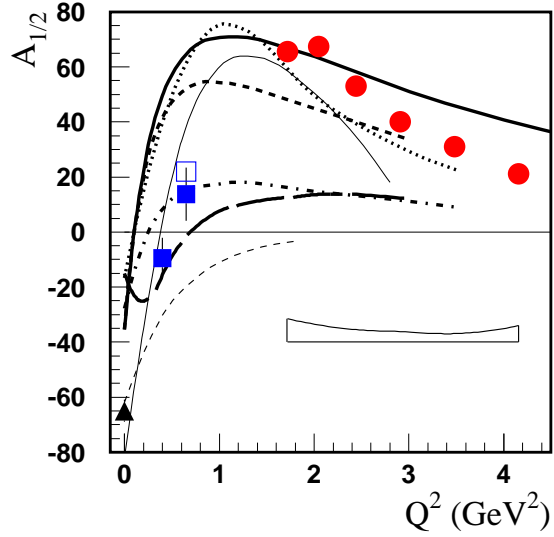}}
\caption{The transverse transition amplitude $A_{1/2}$ in units of $10^{-3} GeV^{-1/2}$ for the Roper resonance, 
clearly showing the change in sign. The black triangle is the PDG average, the full squares are the results of single pion analysis, the open square represents the combined single and double pion analysis. The full circles are preliminary results from CLAS. The curves are predictions of quark model calculations discussed in ref.~\cite{aznauryan2007} with the exception of the thin dashed line which is the prediction of a hybrid baryon model.}
\label{fig:roper}       

\end{figure}

\subsection{The Roper resonance, $P_{11}(1440)$}
\label{sect:3}
The $P_{11}(1440)$ is not a well understood state in the standard CQM. The mass 
is more than 100 MeV lower and the photocoupling amplitude has the wrong sign. 
Alternative models have been developed and make predictions for transition form 
factors, e.g. models using light cone dynamics~\cite{aznauryan2007} 
kinematics, or models describing the state as a hybrid 
baryon~\cite{libuli}. Other models that describe the state as a 
nucleon-meson molecule have been proposed but no transition form factors 
have been computed. The first systematic analyses of the Roper 
transition form factors was accomplished in a combined analysis of $n\pi^+$ and 
$p\pi^0$, and of $p\pi^+\pi^-$ electroproduction data from 
CLAS~\cite{egiyan2006,joo2005,ripani2003} that 
showed a rapid drop of the magnitude of the $A_{1/2}$ amplitude followed 
by a zero-crossing, while the longitudinal coupling $S_{1/2}$ is large 
and positive~\cite{aznauryan2005,aznauryan2005-2}. 

The non-relativistic QCM predicts an incorrect sign at the photon point and
has no zero crossing, the light 
cone quark models give the correct sign at the photon point and predict 
the zero crossing, but lack strength at photon point. This is possibly 
related to contributions of the meson cloud which are not included in 
the light cone (LC) quark model calculations. Meson effects should be less important at 
higher $Q^2$, and better agreement is indeed seen at higher $Q^2$. The analysis 
of new $n\pi^+$ data at high $Q^2$~\cite{park2007,aznauryan2007a} using the unitary isobar 
model (UIM) and dispersion relations (DR) approaches 
result in the behavior shown in Fig.~\ref{fig:roper}. A large 
positive amplitude $A_{1/2}$ is peaking near $Q^2=2$~GeV$^2$, followed 
by a smooth falloff. Both results are quite close and give a consistent behavior, 
indicating that the model-dependence is reasonably well
under control. At large $Q^2$ the $A_{1/2}(Q^2)$ amplitudes drops 
somewhat faster than the LC models predict, which might indicate that the 
point-like coupling to the quarks is not yet realized at these $Q^2$, 
and (constituent) quark form factors are be needed to describe this transition. 

\subsection{The $D_{13}(1520)$ resonance}
\label{sect:4}
\begin{figure}
\resizebox{0.4\textwidth}{!}{%
  \includegraphics{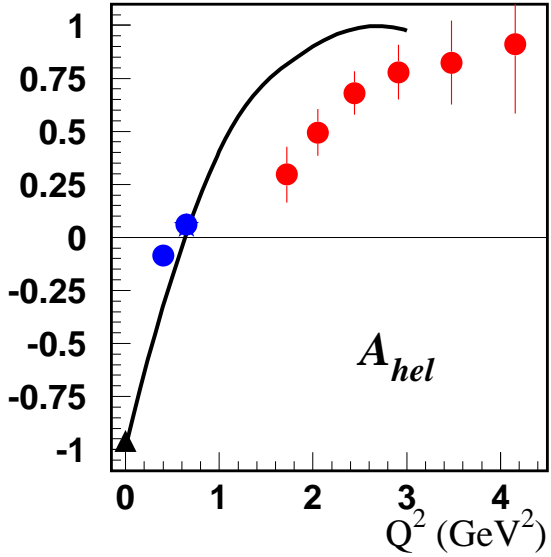}}
\caption{The helicity asymmetry for the $D_{13}(1520)$ state. The full red
symbols are preliminary results from the $n\pi^+$ analysis, while the blue points include $p\pi^0$ and $n\pi^+$ data sets. The curve represents a relativistic quark model calculations ~\cite{aznauryan2008}.}
\label{fig:d13_asym}       
\end{figure}

The $D_{13}(1520)$ is predicted in the CQM to rapidly change its 
helicity structure from helicity 3/2 dominance at the real photon point to helicity 1/2 dominance 
when $Q^2$ increases. Indications of such behavior have been seen 
in previous analyses, but no systematic study has been done in a large $Q^2$ 
range. Figure~\ref{fig:d13_asym} shows the helicity asymmetry 
$$A_{hel} = {{A_{1/2}^2-A_{3/2}^2} \over { A_{1/2}^2+A_{3/2}^2}}$$ extracted from 
the $n\pi^+$ electroproduction data at high $Q^2$. The lower $Q^2$ data 
come from the analysis of $p\pi^0$ and $n\pi^+$ data 
in~\cite{aznauryan2005}. $A_{hel}(Q^2)$ shows the rapid switch in 
helicity dominance. The transition appears to occur in the range $Q^2=0.5-1.0$~GeV$^2$, and the 
asymptotic value is approached at $Q^2>3$~GeV$^2$. 

\section{New photocoupling amplitudes from $\pi^0$ data analysis in full 
resonance region.}
\label{sect:6}
New $\pi^0$ photoproduction data from CLAS have just 
been published \cite{dugger2007} that cover a large angle and energy 
range with high statistics. The SAID analysis pacakge was used to 
determine new photocoupling amplitudes from these data. The 
$S_{11}(1535)$ amplitude determined from the $p\pi^0$ data set now 
agrees very well with the analysis of $p\eta$ data. This result is also 
consistent with the agreement found between these two channels in low 
$Q^2$ electroproduction~\cite{aznauryan2005}, and will hopefully lead to a 
revision of the large uncertainties given in the Review of Particle 
Properties (RPP) for the 
$S_{11}(1535)$ photocoupling amplitude. Another result of the GWU 
analysis is that the $A_{1/2}$ amplitude for the transition to the 
$P_{13}(1720)$ resonance was found as $A^{GWU}_{1/2}(0) = 96.6 \pm 3.4$, 
while the RPP average is listed as $A^{RPP}_{1/2}(0)=18\pm30$, i.e. 
consistent with zero. The new value of $A_{1/2}$ for the $P_{13}(1720)$ 
is qualitatively consistent with the strong excitation of this state 
found earlier in $p\pi^+\pi^-$ electroproduction data from CLAS 
\cite{ripani2003}. The first precision data on the $p\eta^{\prime}$ 
exclusive channel from CLAS have been published 
recently~\cite{dugger2007-1} in the hadronic invariant mass range from 
W=1.95 - 2.25 GeV, and cover the mass range of "missing 
baryons". While there are no clear signals of new s-channel resonances, 
evidence for contributions from the high energy tails of $S_{11}(1535)$ 
and $P_{11}(1710)$ are seen in the data. The analysis of Nakayama and 
Haberzettl~\cite{NH2006} also shows sensitivity to higher mass candidate 
states $P_{11}(2100)$ and $D_{13}(2080)$, that may contribute to a predicted bump 
structure in the total cross section near an invariant mass of 2.09 GeV.

\begin{figure}
\resizebox{0.45\textwidth}{!}{%
  \includegraphics{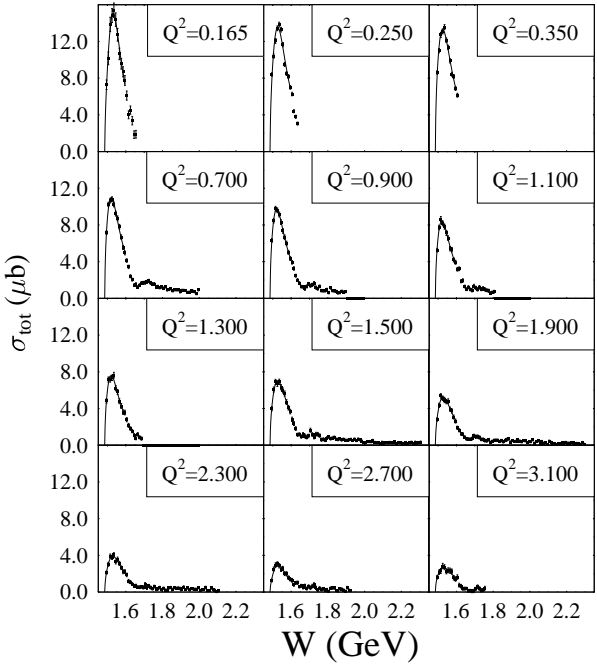}}
\caption{The integrated cross section for $\eta$ production at varies $Q^2$. The large peak is
due to the $S_{11}(1535)$ resonance. In the mass range 1.65  to 1.7 GeV a small dip followed by a peak appears indicating a s-p interference of amplitudes from neighboring resonances.}
\label{fig:SPInter}       
\end{figure}
\section{Search for other excited baryon states.}
\label{sect:7}
A major focus of the CLAS effort is dedicated to clarifying some of the 
ambiguous signals of baryon states, and to the search for new states 
that are predicted within the $SU(6)\otimes O(3)$ symmetry group of 
the symmetric 3-quark 
system. While there are states predicted that represent non-quark 
degrees of freedom, it is important to systematically search for 
predicted 3-quark states. Other contributions, e.g. gluonic excitations 
(hybrid baryons), and nucleon-meson molecule type states will complicate 
the picture, and may require special measurements and analyses 
approaches to separate them from the 3-quark states. The search with 
CLAS aims at complete or nearly complete measurements of a number of 
final states and using linearly and circularly polarized photon beams, 
in combination with longitudinally and transversely polarized targets.

\begin{figure}
\resizebox{0.45\textwidth}{!}{%
  \includegraphics{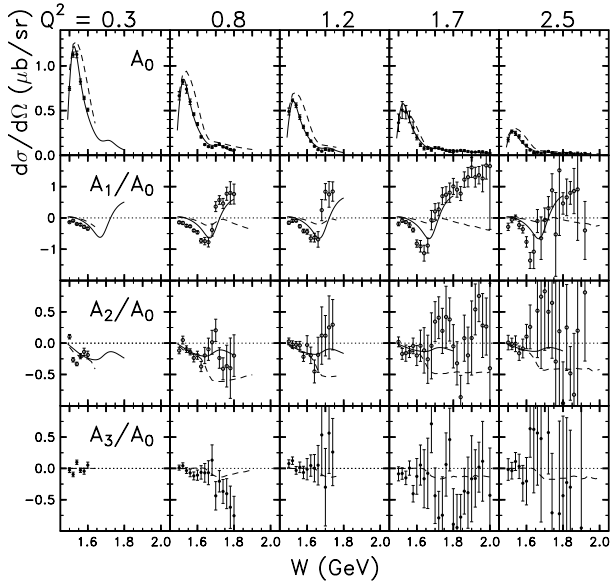}}
\caption{Legendre coefficients $A_0$, and ratios of higher partial wave terms 
for $\eta$ electroproduction data from CLAS plotted versus hadronic mass W. The most 
significant is the ratio $A_1/A_0$ that clearly shows an interference of s- and p-waves 
where one of the waves goes through resonance generating a zero-crossing.}    
\label{fig:A1A0}       
\end{figure}
\subsection{A new P-wave resonance?}
\label{sect:8}
The CLAS collaboration has recently published data on $\eta$ electroproduction
in the mass range from threshold to 2.2 GeV~\cite{denizli2007}. The integrated 
cross section shows a small peak structure near W=1.7 GeV and a dip 
near W=1.65 GeV. This pattern is shown in Fig.~\ref{fig:SPInter} and appears
at all $Q^2$.  To better understand this behavior we expand the response functions in a Legendre polynomial series:
$${d\sigma_T \over d\Omega_{\eta}} + \epsilon {d\sigma_L \over d\Omega_{\eta}} 
= \sum_{l=0}^{\infty} {A_l P_l(\cos\theta^*_{\eta})}$$   
In lowest order the ratio $A_0 / A_1$ can be expressed in terms of the multipoles 
$E_{0+}$ and $M_{1-}$ corresponding to s- and p-waves only, and reads
$${A_1 \over A_0} = {2 {Re}(E^*_{0+}M_{1-})\over {|E_{0+}|^2 + |M_{1-}|^2}}$$
Figure~\ref{fig:A1A0} shows the energy dependence of the ratio $A_1/A_0$. 
It changes sign near $W=1.65$~GeV. The observation is consistent with a rapid 
change in the relative phase of the 
$E_{0+}$ and $M_{1-}$  multipoles because one of them is passing through resonance. 
A reasonable fit to the CLAS data in that mass range is obtained with the $S_{11}(1535)$, 
$S_{11}(1650)$, $P_{11}(1710)$ and $D_{13}(1520)$, with a width for the 
$P_{11}(1710)$ of 100 MeV.
Similar structures, even more pronounced have been observed in $\eta$ photoproduction 
off neutrons, and have been discussed at this 
conference~\cite{shimizu,kuznetsov} as a possible new resonance. Could the observed 
structure be a new resonance? I think it is more likely, that the new data will 
merely confirm the existence of the 3-star $P_{11}(1710)$ state, and better define 
its poorly determined properties such as mass, width, and photocoupling.
\begin{figure}
\resizebox{0.45\textwidth}{!}{%
  \includegraphics{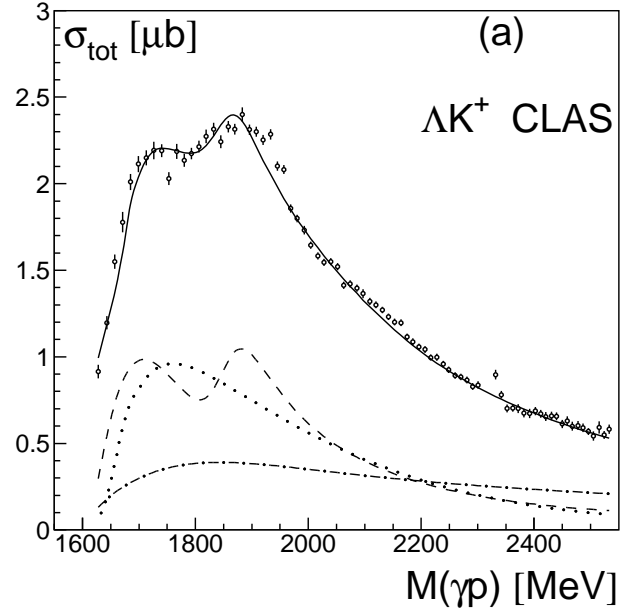}}
\caption{Integrated cross section for $\gamma p \rightarrow K^+\Lambda$. The curves 
are from the Bonn-Gatchina analysis and show contributions from the $P_{11}(1710)$ (dashed-dotted), the $P_{13}(1900)$ (dashed), and from non-resonant K-exchange contributions (dotted).} 
\label{fig:hyperon1}       
\end{figure}

\subsection{Photo- and electroproduction of K-Hyperons}
\label{sect:9}
A large number of cross section data on $K\Lambda$ and $K\Sigma$ 
production have been published and are now being used by various groups 
for phenomenological analyses. The integrated cross section from 
CLAS~\cite{mcnabb2004,bradford2006} is shown in Fig. \ref{fig:hyperon1}. 
These data were used in a fit by the Bonn-Gatchina 
group~\cite{nikonov,anisovich} who found significant 
contributions from the $P_{13}(1900)$, a two star state candidate in the 
RPP. The strongest constraints come from the polarization transfer data 
using a circularly polarized photon beam~\cite{bradford2007}. If the existence 
of the state can be confirmed, it will be strong evidence against a 
diquark-quark model that has no room for such a state~\cite{santopinto}. 
Also, unpolarized and polarized response functions have been measured~\cite{carman,pawel}
 in $\vec{e}p\rightarrow eK^+\Lambda$ and $\vec{e}p \rightarrow eK^+\Sigma$ that show 
significant structures in the hadronic mass spectrum, which are indicative of resonance excitations. 

\begin{figure}
\resizebox{0.5\textwidth}{!}{%
  \includegraphics{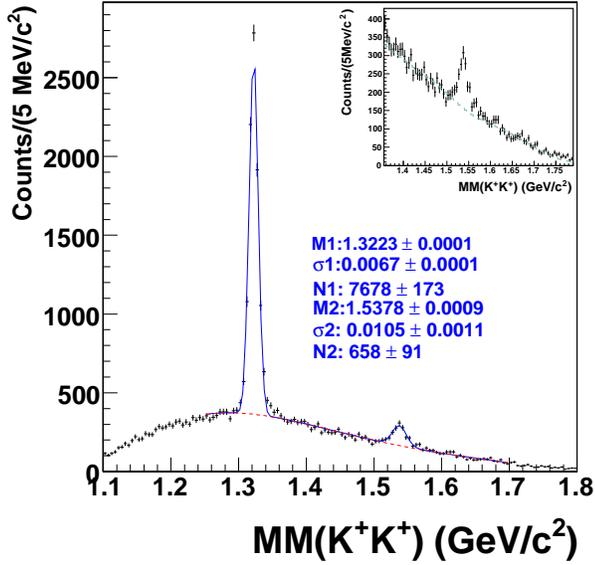}}
\caption{Missing mass for the $\gamma p \rightarrow K^+K^+X^-$ showing the ground state $\Xi^-(1320)$ and the first excited state $\Xi^-(1530)$. The inset highlights the mass range around the 1530 MeV state.}
\label{fig:misskk-cascade}       
\end{figure}
\begin{figure}
\resizebox{0.5\textwidth}{!}{%
  \includegraphics{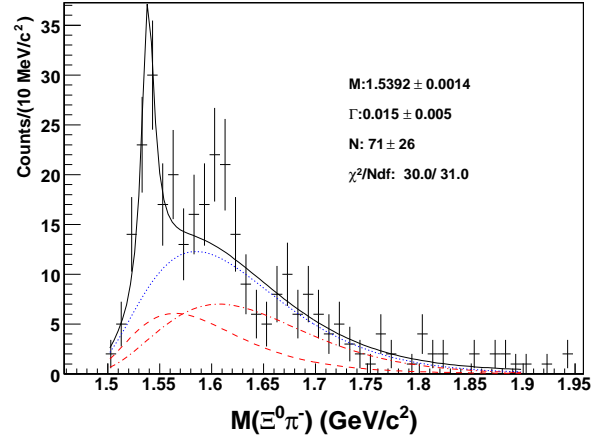}}
\caption{The mass spectrum for $\Xi(1320 \pi^-$ in the reaction $\gamma p \rightarrow \pi^- K^+ K^+ \Xi(1320)$.}
\label{fig:cascade}       
\end{figure}
\subsection{Search for new $\Xi^*$ cascade baryons.}
\label{sect:10}
Production of $\Xi^*$ cascade baryons with strangeness S = -2 provides 
another avenue in the search for new baryon states. The cascade spectrum 
should reflect the same mass splitting due to spin-spin interaction as the S=0 states. 
The advantages are due to the expected (and observed) 
more narrow widths of theses states. The disadvantages for using photon 
beams are the low cross section for the production of two kaons in the 
final state. A possible production 
mechanism is through t-channel production of $K^+\Lambda^*$, where the 
excited $\Lambda^*$ decays through $\Lambda^* \rightarrow K^+\Xi^*$. 
Figure~\ref{fig:misskk-cascade} from CLAS~\cite{guo2007} shows 
that one can identify the lowest two cascade states using the missing mass technique. 
At higher energies other states may become visible as well. Another way to search for 
$\Xi^*$ states is by measuring the $\Xi^0$ with an additional pion.
Forming the 
invariant mass of the $\Xi^0(1320)$ with the $\pi^-$ shows in Fig.~\ref{fig:cascade} 
again the $\Xi(1530)^-$ state. No other structure is clearly identified. Should a 
state at 1.62 GeV emerge at higher statistics, it could be the one star 
candidate in RPP. Such a state would however not be part of the 3-quark symmetry 
group but could be a dynamically generated  $\Xi-\pi$ state prediced in  
dynamical models~\cite{ramos}.
\begin{figure}
\resizebox{0.47\textwidth}{!}{%
  \includegraphics{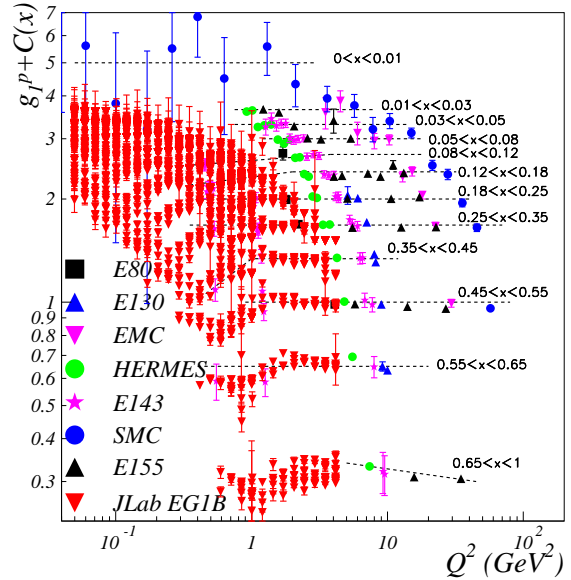}}
\caption{The world data on structure function $g_{1p}(x,Q^2)$. The CLAS data are shown
in the full red triangles.}  
\label{fig:g1p-world}       
\end{figure}
\section{Spin structure of the nucleon and parton distributions}
\label{sect:11}
\begin{figure}
\resizebox{0.5\textwidth}{!}{%
  \includegraphics{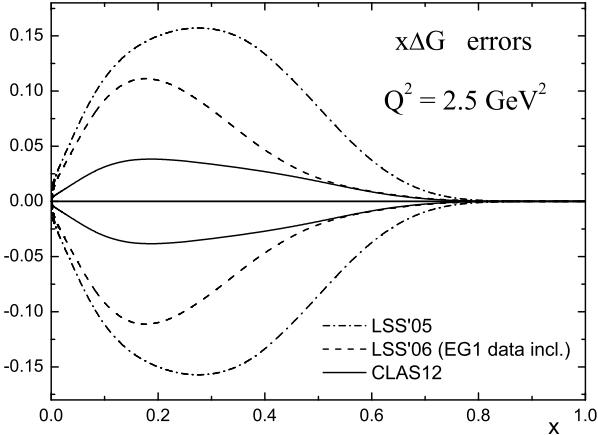}}
\caption{Impact of the CLAS data on the uncertainties in the in the parton 
distribution functions from the LSS QCD analysis. The uncertainty in the 
polarized gluon distribution is reduced by a factor of 3 at a modest $x_B=0.4$ (change from dashed-dotted to dashed lines) giving new constraints on the polarized gluon distribution. The uncertainties in 
the sea quark distribution functions are also improved significantly. The improvement in
the polarized gluon distribution functions comes largely from $g_{1d}(x,Q^2)$ measured
on deuterium in the same kinematics range. 
The projected impact of an extension of the measurements at 12 GeV with the planned 
CLAS12 spectrometer are shown with the solid lines.}  
\label{fig:pdf}       
\end{figure}

The CLAS collaboration has collected very precise data on inclusive double 
polarization inclusive scattering resulting in high quality spin structure 
function $g_{1p}(x,Q^2)$ and $g_{1d}(x,Q^2)$, as well as first moments 
$\Gamma_{1} =\int_{x_{min}}^1g_1(x,Q^2)dx$ for proton, deuterons 
and neutrons. The world data on structure function $g_{1p}(x,Q^2)$ are 
shown in Fig.~\ref{fig:g1p-world}. The CLAS data cover the lower $Q^2$ and 
high $x_B$ range. The bulk of the data covers the resonance region, however the precise 
data in the DIS region provide strong constraints on QCD fits to extract parton 
distribution functions after higher twist contributions have been 
properly taken into account~\cite{LSS2007}. The extracted uncertainties for the polarized 
gluon distribution function are shown in Fig.~\ref{fig:pdf}, and indicate very 
significant reductions compared to results obtained before the CLAS data became available.

\section{Generalized Parton Distributions and DVCS}
\label{sect:12}
\begin{figure}
\resizebox{0.5\textwidth}{!}{%
  \includegraphics{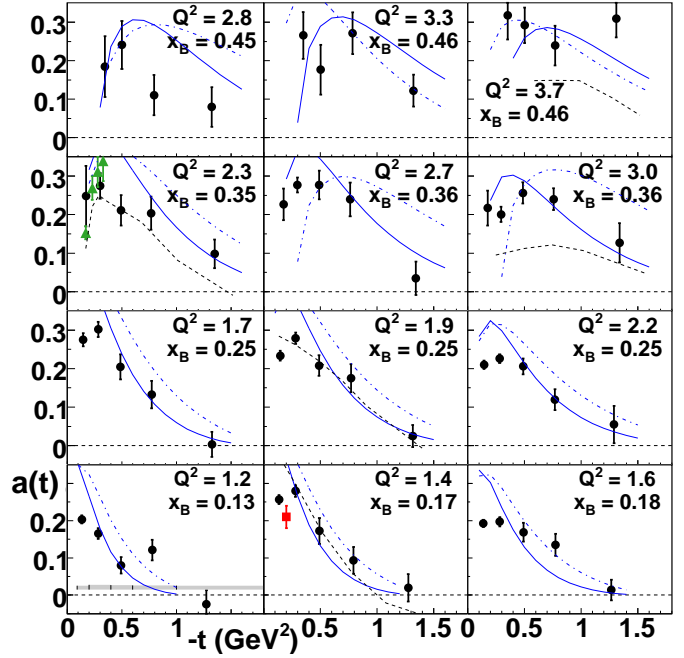}}
\caption{The beam spin asymmetry showing the DVCS-BH interference. The red and green points represent the previous CLAS and Hall A data, respectively. The blue curve is the VGG GPD parameterization~\cite{vgg} in twist-2 (solid) and twist-3 (dashed-dotted).  The dashed black line is a Regge model prediction~\cite{laget}.}
\label{fig:dvcs-asym}       
\end{figure}

The nucleon matrix element of the energy-momentum tensor 
contains 3 form factors that encode information on the angular 
momentum distribution of quark $q$ in transverse space, $J^q(t)$, the 
mass-energy distribution, $M_2^q(t)$, and the pressure and force 
distribution, $d^q_1(t)$. For decades these form factors were of little 
interest as the only known process how they could be directly measured 
is elastic scattering of gravitons off the nucleon. Today we know that 
these form factors also appear as moments of the unpolarized GPDs~\cite{goeke2007}. The 
quark angular momentum in the nucleon is given by $J^q(t) = 
\int_{-1}^{+1}dx[x H^q(x, \xi, t) + E^q(x, \xi, t)]$, which at $t=0$ 
results in the well known Ji sum rule, and $M_2^q(t) + 4/5d^q_1(t)\xi^2 
= \int_{-1}^{+1}dx x H^q(x, \xi, t)$. The mass and pressure distribution 
of the quarks are given by the second moment of GPD $\it{H}$, where the 
latter is probed by parameter $\xi$. A separation of $M^q_2(t)$ and 
$d^q_1(t)$ requires measurement of the moments in a large range of 
$\xi$. How do we access this information? The beam spin asymmetry of the deeply virtual Compton scattering ( DVCS) amplitude interfering with 
the Bethe-Heitler (BH) amplitude is sensitive to the GPD $H(x=\xi,\xi,t)$, and 
has been measured at Jefferson Lab~\cite{clas-dvcs-1,clas-dvcs-2,clas-dvcs-3,halla-dvcs} 
in a wide kinematics range in $Q^2$, $\xi$, and $t$. 
The recent azimuthal asymmetries measured by CLAS were fitted with 
$A_{LU}=\alpha\sin\phi/(1+\beta\cos\phi)$. The $t$-dependence of the leading term 
$\alpha$ for different values of $Q^2$ and $x_B = 2\xi/(1+\xi)$ is 
shown in Fig.~\ref{fig:dvcs-asym}.
We see that $\alpha$ has a maximum at small $t$ and smoothly drops to 
zero. The comparison of $\alpha$ with 
the standard VGG GPD parameterization~\cite{vgg} shows qualitative, even 
quantitative agreement in some kinematics, especially at large $-t$, however the 
theoretical asymmetry exceeds the data at small $-t$. This could mean that at 
low momentum transfer the denominator in the asymmetry does not fully 
account for all contributions to the DVCS cross section. 

While the elastic GPDs are currently at 
the center of the development of a more complex picture of the nucleon, 
GPDs may also be defined for transitions where the recoil baryon is not a ground state 
nucleon but an excited nucleon, such as the $\Delta(1232)$ or any other excited nucleon state. 
Measuring the DVCS with a recoiling excited state allows  probing resonance transitions at the 
parton level, i.e. high $Q^2$, and small momentum transfer $t$, leading to what one may call 
"hard nucleon spectroscopy", a new tool in probing hadronic structures that is not available 
in direct s-channel excitations. 

\section{The Future}
\label{sect:13}
A major focus in current and future experiments is on measurements of polarization observables
in many reaction channels using linearly and circularly polarized photon beams, and longitudinally 
and transverse polarized proton and neutron (deuteron) targets. For strangeness containing channels
often the hyperon recoil polarization can also be measured. Ultimately one would like to obtain a 
model-independent extraction of helicity amplitudes for at least some reaction channels as 
a solid basis in the search for new baryon resonances, and in the determination of  the resonant 
photocoupling amplitudes.  Differential cross sections  including use of circularly polarized photons, 
have already been measured for many processes.  Data taking on 
proton and neutron targets with linearly polarized photons has been 
completed, and the first double and triple polarization observables are 
underway using the polarized proton target FROST. A new frozen spin HD-ice target will be used with 
CLAS in 2010 for measurement of double and triple polarization observables with polarized 
neutrons~\cite{sandorfi}. These data will be extremely  useful in fully coupled channel analyses currently 
under development at EBAC~\cite{harry_lee}.  High statistics DVCS experiments are planned for 2008 and
2009 using polarized electrons and longitudinally polarized target which will provide much more stringent 
constraints on GPDs, and allow determination of some GPDs in specific kinematics. 

\vspace{0.3cm}
\noindent{\bf Ackowledgment}
This work was supported in part by U.S. the Department of Energy and the National Science Foundation, the French commisariat {\'{a}} l'Energie Atomique, the Italian Instituto Nazionale di Fisica Nucleare, the Korea Research Foundation,  and a research grant of the Russian Federation. The Jefferson Science Associates, LLC, operates Jefferson Lab under contract DE-AC05-060R23177.

\end{document}